\RequirePackage[2020-02-02]{latexrelease}
\documentclass[superscriptaddress,nofootinbib,notitlepage]{revtex4-2}
\usepackage[utf8]{inputenc}
\usepackage[T1]{fontenc}
\usepackage{graphicx}
\usepackage{tensor}
\usepackage{booktabs}
\usepackage{array}
\usepackage{paralist}
\usepackage{verbatim}
\usepackage{subfig}
\usepackage{xcolor}
\usepackage{amsfonts}
\usepackage{dcolumn}
\usepackage{amsmath}
\usepackage{amssymb}
\usepackage{bm}
\usepackage{bbm}
\usepackage{hyperref}
\usepackage{fancyhdr}
\usepackage[nottoc,notlof,notlot]{tocbibind}
\usepackage[titles,subfigure]{tocloft}

\begin{document}
\title{Double-extended Kerr-Schild form for $5D$ electrovacuum solutions}

\author{Marcos R. A. Arcod\'ia}
\email{marcodia@iafe.uba.ar}
\affiliation{Instituto de Astronom\'\i
a y F\'\i sica del Espacio (IAFE, CONICET-UBA), Casilla de Correo
67, Sucursal 28, 1428 Buenos Aires, Argentina.}
\affiliation{Departamento de Matem\'atica, Facultad de Ciencias
Exactas y Naturales, Universidad de Buenos Aires, Ciudad
Universitaria, Pabell\'on I, 1428 Buenos Aires,
Argentina.}

\author{Rafael Ferraro}
\email[Member of Carrera del Investigador Cient\'{\i}fico (CONICET,
Argentina); ]{ferraro@iafe.uba.ar}

\affiliation{Instituto de
Astronom\'\i a y F\'\i sica del Espacio (IAFE, CONICET-UBA), Casilla
de Correo 67, Sucursal 28, 1428 Buenos Aires, Argentina.}
\affiliation{Departamento de F\'\i sica, Facultad de Ciencias
Exactas y Naturales, Universidad de Buenos Aires, Ciudad
Universitaria, Pabell\'on I, 1428 Buenos Aires,
Argentina.\vskip.5cm}

\begin{abstract}
Five-dimensional Einstein-Maxwell-Chern-Simons equations are investigated in
the framework of an extended Kerr-Schild strategy to search for black holes
solutions. The fulfillment of Einstein equations constrains the Chern-Simons
coupling constant to a value determined by the trace of the energy-momentum
tensor of the electromagnetic configuration.
\end{abstract}

\maketitle

\vskip1.5cm

\section{Introduction}

The search for Kerr-Newman black holes in higher dimensions has been an
active field of research in the last decades. The $5D$ extremal charged
black hole with a sole angular momentum was obtained in Ref.~\cite{breck}.
This solution does not solve the $5D$ Einstein-Maxwell equations since it
contains Chern-Simons contributions. Numeric \cite{navarro1,navarro2} and
perturbative \cite{mir} methods have also been tried to build this kind of $%
5D$ black hole solutions, but only one angular momentum was considered in
these attempts too.

A family of asymptotically flat rotating solutions to vacuum Einstein
equations in $d$ dimensions was built by Myers and Perry \cite{myersperry}.
These solutions are characterized by $[(d-1)/2]$ independent angular
momenta. Kerr-NUT-(A)dS black holes in higher dimensions have been studied
in Ref.~\cite{knads}.

In five dimensions, the charged rotating black hole with two independent
angular momenta was obtained by Chong, Cveti\v{c}, L\"{u}, and Pope (CCLP)
\cite{cclp}. This geometry appeared in the framework of five-dimensional
minimal gauged supergravity. It is not a solution to Einstein-Maxwell
equations since it is subjected to a specific value of the Chern-Simons (CS)
coupling constant. It is the charged version of the $5D$ Kerr-NUT-(A)dS
black hole. The separability of Hamilton-Jacobi and Klein-Gordon in the CCLP
chart was studied in Ref.~\cite{DKL}. The existence of a Killing-Yano tensor
for this solution is established in Ref.~\cite{KKY}.

Equal angular momenta solutions to Einstein-Maxwell-Chern-Simons (EMCS)
equations with arbitrary values of the CS coupling constant have been
investigated in Ref.~\cite{navarro2,mir}, although from a numerical or
perturbative perspective. The analytical building of solutions to EMCS
equations with arbitrary CS coupling constant and two independent angular
momenta was studied in Ref.~\cite{ferraro}, by introducing a properly
extended Pleba\'{n}ski-Demia\'{n}ski \textit{Ansatz} \cite{PD} in $5D$. In
this work we will renew the search for this kind of solutions in the
framework of a \textit{double-extended Kerr-Schild form} for the metric
tensor. We will describe the flat $5D$ seed geometry by means of a null
geodesic tetrad in a $(2+2)D$ submanifold together with an orthogonal
unitary vector. The chart to be used renders this basis in a simple form,
what will prove to be very friendly for algebraic manipulations. We will
show that the specific CS coupling constant of the CCLP solution seems to be
inescapable, since it is determined by the trace of the energy-momentum
tensor of the pointlike solution to Maxwell-Chern-Simons equation.

\bigskip

In Section \ref{II} we introduce the basic properties of the chart
describing the seed metric. In Section \ref{III} we describe the Kerr-Schild
strategy to search for black-hole solutions, and its different extensions.
In Section \ref{IV} we show the pointlike solution to Maxwell-Chern-Simons
equation in a double-extended Kerr-Schild background. In Section \ref{V} we
solve the EMCS equations for the pointlike charge. In Section \ref{VI} we establish the relationship between signature (3,2) and (1,4). In Section \ref{VII} we
display the conclusions.

\bigskip

\section{From flat spacetime to rotating black holes}

\label{II}

Let be ${\{}\mathbf{n}_{a}\}$ a basis for the tangent space of a $5D$
spacetime ($a=0,...,4$), and $\{{n}^{a~}\}$ the respective dual basis of
1-forms,
\begin{equation}
{n}^{a}(\mathbf{n}_{b})=\delta _{b}^{a}~.  \label{dual}
\end{equation}%
The metric tensor will be formulated as
\begin{equation}
\mathbf{g}={n^{0}}\otimes {n}^{2}+{n}^{2}\otimes {n^{0}}+{n^{1}}\otimes {n}%
^{3}+{n}^{3}\otimes {n^{1}}+{n}^{4}\otimes {n}^{4}~.  \label{metric}
\end{equation}%
Due to the duality (\ref{dual}) the inverse metric results to be
\begin{equation}
\mathbf{g}^{-1}={\mathbf{n}_{2}}\otimes {\mathbf{n}_{0}}+\mathbf{n}%
_{0}\otimes {\mathbf{n}_{2}}+\mathbf{n}_{1}\otimes {\mathbf{n}_{3}}+{\mathbf{%
n}_{3}}\otimes {\mathbf{n}_{1}}+\mathbf{n}_{4}\otimes \mathbf{n}_{4}~.
\label{inverse}
\end{equation}%
The form of the metric implies that vectors $\{{\mathbf{n}_{0}},~{\mathbf{n}%
_{1}},~\mathbf{n}_{2},~{\mathbf{n}_{3}\}}$ are null;\ they form a basis of a
$2+2$ spacetime,
\begin{equation}
|\mathbf{n}_{a}|^{2}=\mathbf{n}_{a}\cdot \mathbf{n}_{a}=\mathbf{g}(\mathbf{n}%
_{a},~\mathbf{n}_{a})=0~,~~~\ ~~~~a=0,~1,~2,~3~.
\end{equation}%
Vector $\mathbf{n}_{4}$ is unitary,%
\begin{equation}
|\mathbf{n}_{4}|^{2}=\mathbf{n}_{4}\cdot \mathbf{n}_{4}=\mathbf{g}(\mathbf{n}%
_{4},~\mathbf{n}_{4})=1~.
\end{equation}%
The metric (\ref{metric}) and the inverse metric (\ref{inverse}) can be
written as%
\begin{equation}
\mathbf{g}=\breve{\eta}_{ab}~{n}^{a}\otimes {n}^{b}~,~\ \ ~\ \ \ \ \ \ ~\
\mathbf{g}^{-1}=\breve{\eta}^{ab}~\mathbf{n}_{a}\otimes \mathbf{n}_{b}~,
\label{gg}
\end{equation}%
where%
\begin{equation}
\breve{\eta}^{ab}=\breve{\eta}_{ab}=%
\begin{pmatrix}
0 & 0 & 1 & 0 & 0 \\
0 & 0 & 0 & 1 & 0 \\
1 & 0 & 0 & 0 & 0 \\
0 & 1 & 0 & 0 & 0 \\
0 & 0 & 0 & 0 & 1%
\end{pmatrix}%
\end{equation}%
is an orthogonal transformation of the $(2+3)D$ Minkowski symbol:%
\begin{equation}
\breve{\eta}_{ab}=\frac{1}{2}~%
\begin{pmatrix}
-1 & 0 & 1 & 0 & 0 \\
0 & 1 & 0 & 1 & 0 \\
1 & 0 & 1 & 0 & 0 \\
0 & 1 & 0 & -1 & 0 \\
0 & 0 & 0 & 0 & \sqrt{2}%
\end{pmatrix}%
\begin{pmatrix}
-1 & 0 & 0 & 0 & 0 \\
0 & 1 & 0 & 0 & 0 \\
0 & 0 & 1 & 0 & 0 \\
0 & 0 & 0 & -1 & 0 \\
0 & 0 & 0 & 0 & 1%
\end{pmatrix}%
\begin{pmatrix}
-1 & 0 & 1 & 0 & 0 \\
0 & 1 & 0 & 1 & 0 \\
1 & 0 & 1 & 0 & 0 \\
0 & 1 & 0 & -1 & 0 \\
0 & 0 & 0 & 0 & \sqrt{2}%
\end{pmatrix}%
\end{equation}

For some purposes it is useful to call%
\begin{equation}
\mathbf{m~}\equiv ~\mathbf{n}_{4}~,
\end{equation}%
and preserve the symbol $\mathbf{n}_{a}$ for the null tetrad ($a=0,~1,~2,~3$%
).

Other properties emerging from the form of the metric (\ref{metric}) are: $%
\mathbf{m}$ is orthogonal to the null tetrad,%
\begin{equation}
\mathbf{m}\cdot \mathbf{n}_{a}=\mathbf{g}(\mathbf{m},\mathbf{n}%
_{a})=0~,~~~~~~~~a=0,~1,~2,~3~,
\end{equation}%
and vectors $\mathbf{n}_{a}$ fulfill the orthogonality relations%
\begin{equation}
\mathbf{n}_{0}\cdot \mathbf{n}_{1}=0=\mathbf{n}_{2}\cdot \mathbf{n}_{3}~,~~~~%
\mathbf{n}_{1}\cdot \mathbf{n}_{2}=0=\mathbf{n}_{0}\cdot \mathbf{n}_{3}~.
\end{equation}%
We remark the way to ``lower the indices'' of the null tetrad, since the
metric (\ref{metric}) mixes the characters:%
\begin{equation}
\mathbf{g}(\mathbf{n}_{2},~~)={n}^{0},~~\ ~\ \mathbf{g}(\mathbf{n}_{3},~~)={n%
}^{1},~\ \ \ \ \ \text{etc.}  \label{lower}
\end{equation}

\bigskip

\subsection{Flat spacetime}

The former framework will be used to search for solutions to
Einstein-Maxwell equations (electrovacuum solutions) in five dimensions. Let
us first introduce the flat metric, which will play the role of a \textit{%
seed} metric in following steps. The flat spacetime can be represented by
choosing a chart $(t,\phi ,\psi ,r,p)$ such that%
\begin{equation}
{n}^{2}=dr~,~~~\ \ \ ~~~{n}^{3}=dp~.  \label{n2n3}
\end{equation}%
The basis of the cotangent space is completed with the 1-forms
\begin{eqnarray}
{n^{0}} &=&dt-\frac{a^{2}+p^{2}}{a}~d\phi -\frac{b^{2}+p^{2}}{b}~d\psi \ ,
\label{n0seed} \\
{n^{1}} &=&dt-\frac{a^{2}+r^{2}}{a}~d\phi -\frac{b^{2}+r^{2}}{b}~d\psi \ ,
\label{n1seed} \\
{m} &=&r~p~\left( -\frac{b}{a^{2}}~d\phi -\frac{a}{b^{2}}~d\psi \right) \ .
\label{mu}
\end{eqnarray}%
Thus the metric exhibits a symmetry under the exchange $r\longleftrightarrow
p$, ${n^{0}}\longleftrightarrow {n^{1}}$. The components of $\mathbf{g}^{%
\text{flat}}$ are%
\begin{equation}
g_{\mu \nu }^{\text{flat}}=%
\begin{pmatrix}
0 & 0 & 0 & 1 & 1 \\
0 & \frac{b^{2}p^{2}r^{2}}{a^{4}} & \frac{p^{2}r^{2}}{ab} & -\frac{%
a^{2}+p^{2}}{a} & -\frac{a^{2}+r^{2}}{a} \\
0 & \frac{p^{2}r^{2}}{ab} & \frac{a^{2}p^{2}r^{2}}{b^{4}} & -\frac{%
b^{2}+p^{2}}{b} & -\frac{b^{2}+r^{2}}{b} \\
1 & -\frac{a^{2}+p^{2}}{a} & -\frac{b^{2}+p^{2}}{b} & 0 & 0 \\
1 & -\frac{a^{2}+r^{2}}{a} & -\frac{b^{2}+r^{2}}{b} & 0 & 0%
\end{pmatrix}%
~,  \label{seedmet}
\end{equation}%
whose determinant is%
\begin{equation}
\det (g_{\mu \nu }^{\text{flat}})~=~\frac{r^{2}p^{2}}{a^{4}b^{4}}%
~(a^{2}-b^{2})^{2}~(r^{2}-p^{2})^{2}~.  \label{det}
\end{equation}

As can be seen, in this \textit{Ansatz} the components of $\mathbf{g}^{\text{%
flat}}$ look much simpler than they are in other \textit{Ans\"{a}tze}
appearing in the literature, what is a great advantage for doing
calculations. The  parameters $a$, $b$  will be related to
the independent angular momenta of the $5D$ black hole solutions. To avoid the
vanishing of $\det (g_{\mu \nu }^{\text{flat}})$, the otherwise arbitrary
parameters $a$, $b$ must be unequal; in fact, the second and third columns
of $\det (g_{\mu \nu }^{\text{flat}})$ would be equal in case that $a=b$.%
\footnote{%
In five dimensions, equal angular momenta implies a static geometry \cite%
{Anabalon}.} The fact that the metric (\ref{seedmet}) describes a flat
spacetime is straightforwardly verifiable by computing the Riemann tensor.
The components of the metric tensor in the chart $(t,\phi ,\psi ,r,p)$ also
show that the Killing vectors $\partial _{\phi }$, $\partial _{\psi }$ have
positive squared norm, while $\partial _{t}$ is null.

\bigskip

Duality (\ref{dual}) implies that%
\begin{eqnarray}
\mathbf{n}_{0} &=&\frac{1}{r^{2}-p^{2}}~\left( \left(
a^{2}+b^{2}+r^{2}\right) ~\frac{\partial }{\partial t}+\frac{a^{3}}{%
a^{2}-b^{2}}~\frac{\partial }{\partial \phi }+\frac{b^{3}}{b^{2}-a^{2}}~%
\frac{\partial }{\partial \psi }\right) \ ,  \label{n0} \\
\mathbf{n}_{1} &=&-\frac{1}{r^{2}-p^{2}}~\left( \left(
a^{2}+b^{2}+p^{2}\right) ~\frac{\partial }{\partial t}+\frac{a^{3}}{%
a^{2}-b^{2}}~\frac{\partial }{\partial \phi }+\frac{b^{3}}{b^{2}-a^{2}}~%
\frac{\partial }{\partial \psi }\right) \ ,  \label{n1} \\
\mathbf{m} &=&\frac{1}{r~p}\left( ab~\frac{\partial }{\partial t}+\frac{%
a^{2}b}{a^{2}-b^{2}}~\frac{\partial }{\partial \phi }+\frac{b^{2}a}{%
b^{2}-a^{2}}~\frac{\partial }{\partial \psi }\right) ~,  \label{m}
\end{eqnarray}%
\begin{equation}
\mathbf{n}_{2}=\frac{\partial }{\partial r}\doteq \mathbf{k~,~~\ \ \ \ \ \ \
~\ ~~n}_{3}=\frac{\partial }{\partial p}\doteq \mathbf{K~.}
\label{kKvectors}
\end{equation}%
According to Eq.~(\ref{lower}), by lowering indices to the vectors $\mathbf{%
k~,K}$ one gets%
\begin{equation}
\mathbf{g}(\mathbf{k},~~)={n}^{0}\doteq {k}~,~~\ ~\ \mathbf{g}(\mathbf{K}%
,~~)={n}^{1}\doteq {K}~.~  \label{kKforms}
\end{equation}

\bigskip

\subsection{Ansatz to search for black hole solutions}

To search for black hole solutions we will introduce four functions $X,Y,{%
\mathcal{X}},{\mathcal{Y}}$ to be determined by solving the Einstein-Maxwell
equations. These unknown functions enter the metric through the replacements

\begin{equation}
dr\longrightarrow {\underline{n}}^{2}\doteq dr{-}\frac{Y(r)}{2~(r^{2}-p^{2})}%
{~k}\ ,~~\ ~~\ \ dp\longrightarrow {\underline{n}}^{3}\doteq dp{+}\frac{X(p)%
}{2~(r^{2}-p^{2})}{~K}\ ,~~\ ~~\ \ {m}\longrightarrow {\underline{n}}%
^{4}\doteq {m}+\frac{-\mathcal{Y}(r)~{k}+\mathcal{X}(p)~{K}}{%
~r^{-1}p^{-1}(r^{2}-p^{2})}~,  \label{ansatz1}
\end{equation}

\bigskip

\noindent where ${k}\ ,{K}$ are the (null) 1-forms (\ref{kKforms}). To keep
the duality of bases we must also change

\begin{equation}
\mathbf{n}_{0}\longrightarrow {\mathbf{\bar{n}}_{0}\doteq \mathbf{n}_{0}+}%
\frac{Y(r)}{2~(r^{2}-p^{2})}{~}\mathbf{k}{+}\frac{r~p~{\mathcal{Y}(}r{)}}{%
r^{2}-p^{2}}~{\mathbf{m}}\ \ ,~~\ ~~~~~\ \ \mathbf{n}_{1}\longrightarrow {%
\mathbf{\bar{n}}_{1}\doteq \mathbf{n}_{1}-}\frac{X(p)}{2~(r^{2}-p^{2})}{~}%
\mathbf{K}{-}\frac{r~p~{\mathcal{X}(}p{)}}{r^{2}-p^{2}}~{\mathbf{m}}~,
\label{ansatz2}
\end{equation}

\bigskip

\noindent where $\mathbf{k,K}$ are the null vectors (\ref{kKvectors}). This
strategy to introduce gravity starting from a seed flat metric makes contact
with both the Kerr-Schild \cite{Kerr65,Debney67,Kerr07} and the Pleba\'{n}%
ski-Demia\'{n}ski \cite{PD,Podolsky} \textit{Ans\"{a}tze}, which were
originally designed to try with $4D$ solutions. Since the \textit{Ansatz} (%
\ref{ansatz1}, \ref{ansatz2}) does not alter the 5-form of volume ${%
n^{0}\wedge n^{1}\wedge n^{2}}\wedge {n^{3}}~{\wedge ~n^{4}}$ (${=k\wedge
K\wedge }dr\wedge dp~{\wedge ~m}$), the determinant (\ref{det}) does not
change.

\medskip

\subsection{Null geodesics}

For a null vector $k^{\mu }$ to be geodesic it must satisfy the equation%
\footnote{%
A geodesic null vector $\mathbf{k}$ fulfills $k_{\mu }k^{\mu }=0$ and $%
k^{\mu }k_{\nu ;\mu }=0$; then it is $0=k^{\mu }(k_{\nu ;\mu }-k_{\mu ;\nu
})=k^{\mu }(k_{\nu ,\mu }-k_{\mu ,\nu })$.}%
\begin{equation}
k^{\mu }~(\partial _{\mu }k_{\nu }-\partial _{\nu }k_{\mu })=0~~\ \ \ \ \ \ ~%
\text{or}~~\ \ \ \ \ \ ~d{k}(\mathbf{k},~)=0~.
\end{equation}%
It is easy to verify that $d{k}(\mathbf{k},~)$ (i.e., $d{n^{0}}(\frac{%
\partial }{\partial r},~)~$) and $d{K}(\mathbf{K},~)$ (i.e., $d{n^{1}}(\frac{%
\partial }{\partial p},~)~$) are zero for the original bases. On the other
hand, the \textit{Ansatz} (\ref{ansatz1}, \ref{ansatz2}) does not modify the
1-forms ${n^{0}}={k}$ and ${n^{1}}={K}$, nor the null vectors $\mathbf{n}%
_{2}=\mathbf{k}$ and$\mathbf{~n}_{3}=\mathbf{K}$. Thus, $\mathbf{k},\mathbf{K%
}$ remain null and geodesic whatever the unknown functions $X,Y,{\mathcal{X}}%
,{\mathcal{Y}}$ are.\footnote{%
On the contrary, the null vectors $\mathbf{n}_{0},\mathbf{n}_{1}$ are
geodesic only in the seed flat spacetime we began with.}

For a later use we will mention two other equations that are independent of
the functions $X,Y,{\mathcal{X}},{\mathcal{Y}}$. Since vector $\mathbf{m}$
is not modified by the \textit{Ansatz}, it also remains valid that
\begin{equation}
m^{\mu }~(\partial _{\mu }k_{\nu }-\partial _{\nu }k_{\mu })=0~,~~\ \ ~\ \
~~~~~m^{\mu }~(\partial _{\mu }K_{\nu }-\partial _{\nu }K_{\mu })=0~.
\label{dkdK}
\end{equation}

\medskip

\subsection{Kerr-NUT-(A)dS geometry}

Remarkably, the metric $\mathbf{g}$ is flat not only when functions $X,Y,{%
\mathcal{X}},{\mathcal{Y}}$ are zero.\ It remains flat if

\begin{equation}
X_{\text{flat}}(p)=\gamma +\alpha ~p^{2}-\beta ^{2}~p^{-2},~~~\ Y_{\text{flat%
}}(r)=\gamma +\alpha ~r^{2}-\beta ^{2}~r^{-2},~~\ {\mathcal{X}}_{\text{flat}%
}(p)=\delta +\varepsilon ~p^{2}+\beta ~p^{-2},~~{\mathcal{Y}}_{\text{flat}%
}(r)=\delta +\varepsilon ~r^{2}+\beta ~r^{-2},  \label{flatfunctions}
\end{equation}

\bigskip

\noindent what means that the free coefficients $\alpha ,\beta ,\gamma
,\delta ,\varepsilon $ entail changes of coordinates. The \textit{Ansatz} (%
\ref{ansatz2}) allows to easily introduce the cosmological constant $\Lambda
$, the mass $m$ and the NUT charge $n$; the $5D$ Kerr-NUT-(A)dS geometry is
given by%
\begin{equation}
X(p)=X_{\text{flat}}(p)+2n-\frac{\Lambda }{6}~p^{4}~,~~~\ ~\ \ Y(r)=Y_{\text{%
flat}}(r)-2m-\frac{\Lambda }{6}~r^{4}~,\ \   \label{KerrNUT}
\end{equation}%
as can be verified by solving the vacuum Einstein equations with
cosmological constant \cite{GLPP}. At this level, the solution acquires two
independent angular momenta characterized by the values of $a,b$.  Notice that
the case $m=-n$ is not a black hole but
is de Sitter spacetime, since $m,n$ would be absorbed into a (single)
constant $\gamma $ in $X_{\text{flat~}}$, $Y_{\text{flat~}}$. As seen, the $%
5D$ mass and NUT charge are somehow degenerate since one of them can be
absorbed into the other one by shifting the constant $\gamma $ \cite{CGLP}.
It could be said that the $5D$ mass and NUT charge are introduced by
replacing the single constant $\gamma $ in Eq.~(\ref{flatfunctions}) with
two different constants. Analogously, the electric charge will be introduced
in Section \ref{IV} by replacing the single constant $\beta $ in ${\mathcal{X%
}}_{\text{flat}}$, ${\mathcal{Y}}_{\text{flat}}$ with two different
constants.

\bigskip

\section{The double-extended Kerr-Schild form}

\label{III}

By replacing the \textit{Ansatz} (\ref{ansatz2}) in the inverse metric (\ref%
{inverse}) we obtain a structure of type \emph{double-extended Kerr-Schild }(%
\emph{dxKS}),
\begin{equation}
\mathbf{g}^{-1}=\mathbf{h}^{-1}+\mathcal{H}_{0}~\mathbf{k}\otimes \mathbf{k}+%
\mathcal{H}_{1}~\mathbf{K}\otimes \mathbf{K}+\mathcal{J}_{0}~(\mathbf{k}%
\otimes \mathbf{m}+\mathbf{m}\otimes \mathbf{k})+\mathcal{J}_{1}~(\mathbf{K}%
\otimes \mathbf{m}+\mathbf{m}\otimes \mathbf{K})~,  \label{dxKSinverse}
\end{equation}%
where $\mathbf{h}$ is the seed metric, $\mathbf{k}$ and $\mathbf{K}$ are
mutually orthogonal null geodesic vector fields with respect to both $%
\mathbf{g}$ and $\mathbf{h}$, and $\mathbf{m}$ is a unitary vector
orthogonal to both $\mathbf{k}$ and $\mathbf{K}$. In the \textit{Ansatz} (%
\ref{ansatz2}) the functions $\mathcal{H}$, $\mathcal{J}$ are
\begin{equation}
\mathcal{H}_{0}=\frac{Y(r)}{r^{2}-p^{2}}~,\ \ \ \ \ \ \mathcal{H}_{1}=-\frac{%
X(p)}{r^{2}-p^{2}}~,\ \ \ \ \ \ \mathcal{J}_{0}=\frac{r\ p\ \mathcal{Y}(r)}{%
r^{2}-p^{2}}~,\ \ \ \ \ \ \mathcal{J}_{1}=-\frac{r\ p\ \mathcal{X}(p)}{%
r^{2}-p^{2}}~,  \label{dxKSansatz}
\end{equation}%
and the seed metric $\mathbf{h}$ is the one in Eq.~(\ref{metric}). Of
course, we can get rid of functions $\mathcal{J}$ in the $5D$ Kerr-NUT-(A)dS
solution, since the free constants $\delta $, $\varepsilon $, $\beta $ can
be chosen to be zero. Even the cosmological term in Eq.~(\ref{KerrNUT})
could be hidden in $\mathbf{h}^{-1}$ to cast the Kerr-(A)dS solution into
the simpler $KS$ form \cite{Kerr65,Debney67,Kerr07} where $\mathcal{H}_{1}$,
$\mathcal{J}_{0}$, $\mathcal{J}_{1}$ are zero. The $KS$ form is connected to
the Newman-Janis algorithm to get the rotating charged black hole in four
dimensions \cite{Ferraro14}.

\medskip

The case of the \emph{extended Kerr-Schild} form (\emph{xKS})
\begin{equation}
\mathbf{g}^{-1}=\mathbf{h}^{-1}+\mathcal{H~}\mathbf{k}\otimes \mathbf{k}+%
\mathcal{J~}(\mathbf{k}\otimes \mathbf{m}+\mathbf{m}\otimes \mathbf{k})~,
\label{xKSinverse}
\end{equation}%
has been studied as a way of looking for solutions in higher dimensions \cite%
{Malek}, in particular as a form of containing the CCLP solution \cite%
{Aliev, xKS}. However, the $dxKS$ form provide us with a wider menu of
possibilities to succeed in searching for new electrovacuum solutions.

The metric $\mathbf{g}$ related to $\mathbf{g}^{-1}$ in Eq.~(\ref%
{dxKSinverse}) is
\begin{equation}
\mathbf{g}=\mathbf{h}+\left( \mathcal{J}_{0}^{~2}-\mathcal{H}_{0}\right) ~{k}%
\otimes {k}+\left( {\mathcal{J}_{1}}^{2}-\mathcal{H}_{1}\right) ~{K}\otimes {%
K}+\mathcal{J}_{0}\mathcal{J}_{1}~({k}\otimes {K}+{K}\otimes {k})-\mathcal{J}%
_{0}~({k}\otimes {m}+{m}\otimes {k})-\mathcal{J}_{1}~({K}\otimes {m}+{m}%
\otimes {K})~.  \label{dxKSg}
\end{equation}%
For $\mathbf{g}^{-1}$ in Eq.~(\ref{dxKSinverse}) to be the inverse of $%
\mathbf{g}$, the 1-forms $k,~K,~{m}$ in Eq.~(\ref{dxKSg}) must be defined as%
\begin{equation}
k_{\mu }\doteq h_{\mu \nu }~k^{\nu }=g_{\mu \nu }~k^{\nu },~~\ ~\quad K_{\mu
}\doteq h_{\mu \nu }~K^{\nu }=g_{\mu \nu }~K^{\nu },~~\ ~\quad m_{\mu
}\doteq h_{\mu \nu }~m^{\nu }
\end{equation}%
Thus, the proof is achieved by using that $k_{\mu }k^{\mu }=K_{\mu }K^{\mu
}=k_{\mu }K^{\mu }=k_{\mu }m^{\mu }=K_{\mu }m^{\mu }=0$, and $h_{\mu \nu
}~m^{\mu }m^{\nu }=1$. Notice that $m_{\mu }$ does not coincide with $g_{\mu
\nu }~m^{\nu }$:%
\begin{equation}
g_{\mu \nu }~m^{\nu }=h_{\mu \nu }~m^{\nu }-\mathcal{J}_{0}~k_{\mu }-%
\mathcal{J}_{1}~K_{\mu }=m_{\mu }-\mathcal{J}_{0}~k_{\mu }-\mathcal{J}%
_{1}~K_{\mu }~.
\end{equation}%
This is not an obstacle for vector $\mathbf{m}$ to be unitary in both
metrics $\mathbf{h}$ and $\mathbf{g}$, since $\mathbf{m}$ is orthogonal to $%
\mathbf{k}$ and $\mathbf{K}$.

\bigskip

\section{Electrodynamics in the $dxKS$ Ansatz}

\label{IV}

Our aim is to obtain $5D$ charged rotating black holes, so we will now turn
the attention to Maxwell equations in the $dxKS$ framework. The
\textquotedblleft pointlike\textquotedblright\ charge potential\footnote{%
Both terms are gauge equivalents since $(r^{2}-p^{2})^{-1}~k$ and $%
(r^{2}-p^{2})^{-1}~K$ differs in an exact 1-form.}%
\begin{equation}
{A}=\frac{Q}{2~(r^{2}-p^{2})}~\left( {n^{0}}+{n^{1}}\right)
\label{potential}
\end{equation}%
yields a solution to Maxwell equations
\begin{equation}
d\ast {F}=0  \label{Maxwell}
\end{equation}%
not only in the seed flat metric but for arbitrary functions $X,Y$, as will
be shown. In fact, the corresponding electromagnetic field is \footnote{%
According to Eq.~(\ref{ansatz1}) it is ${n^{0}}\wedge {\underline{n}}^{2}={%
n^{0}}\wedge n^{2}$, ${n^{1}}\wedge {\underline{n}}^{3}={n^{1}}\wedge n^{3}$
(remember that $k=n^{0}$ and $K=n^{1}$).}
\begin{equation}
{F}=d{A}=\frac{2Q~r}{\left( r^{2}-p^{2}\right) ^{2}}~{n^{0}}\wedge {n}^{2}-%
\frac{2Q~p}{\left( r^{2}-p^{2}\right) ^{2}}~{n^{1}}\wedge {n}^{3}~,
\label{field}
\end{equation}%
\begin{equation}
\ast {F}=\frac{2Q~r}{\left( r^{2}-p^{2}\right) ^{2}}~{n^{1}}\wedge {n}%
^{3}\wedge \underline{n}^{4}-\frac{2Q~p}{\left( r^{2}-p^{2}\right) ^{2}}~{%
n^{0}}\wedge {n}^{2}\wedge {\underline{n}}^{4}~.  \label{dualfield}
\end{equation}%
No trace of $X,Y$ is left in ${F}$ and $\ast {F}$. However $\ast {F}$ is
affected by the functions ${\mathcal{X}},{\mathcal{Y}}$, which enter $\ast {F%
}$ through $\underline{n}^{4}$. The field (\ref{field}) no longer fulfills
Maxwell equations (\ref{Maxwell}) in a general $dxKS$ \textit{Ansatz}
because $d\ast {F}$ will depend on the functions ${\mathcal{X}},{\mathcal{Y}}
$ (or $\mathcal{J}_{0},\mathcal{J}_{1}$) and their derivatives. Certain
choices of these functions could eventually restore a proper dynamical
behavior for the potential (\ref{potential}).
\begin{equation*}
d\ast {F}=d\ast {\mathring{F}}+d\left( \frac{2Q~r}{\left( r^{2}-p^{2}\right)
^{2}}~{n^{1}}\wedge {n}^{3}\wedge \frac{-\mathcal{Y}(r)~{n^{0}}+\mathcal{X}%
(p)~{n^{1}}}{~r^{-1}~p^{-1}~(r^{2}-p^{2})}-\frac{2Q~p}{\left(
r^{2}-p^{2}\right) ^{2}}~{n^{0}}\wedge {n}^{2}\wedge \frac{-\mathcal{Y}(r)~{%
n^{0}}+\mathcal{X}(p)~{n^{1}}}{~r^{-1}~p^{-1}~(r^{2}-p^{2})}\right) ~,
\end{equation*}%
where $\ast {\mathring{F}}$ stands for the dual field in the seed metric,
whose exterior derivative vanishes for the potential (\ref{potential}). Then
the former expression reduces to
\begin{equation}
d\ast {F}=d\left( -\frac{2Q~p~r^{2}~\mathcal{Y}(r)~}{\left(
r^{2}-p^{2}\right) ^{3}}~{n^{0}\wedge n^{1}}\wedge {\underline{n}}^{3}+\frac{%
2Q~r~p^{2}~\mathcal{X}(p)}{\left( r^{2}-p^{2}\right) ^{3}}~{n^{0}}\wedge {%
n^{1}}\wedge {\underline{n}}^{2}\right) ~.  \label{dsF}
\end{equation}%
Instead of a direct computation of the exact 4-form in the second term, we
can resort to the relations
\begin{equation}
d{n}^{a}=\frac{1}{2}~c_{~cb}^{a}~{n}^{b}\wedge {n}^{c}  \label{dn}
\end{equation}%
where $c_{~cb}^{a}$ are the structure functions defined by
\begin{equation}
\lbrack \mathbf{n}_{c}~,~\mathbf{n}_{b}]=c_{~cb}^{a}~\mathbf{n}_{a}~.
\end{equation}%
(see for instance equation (6.176) in Ref.~\cite{gron}). In the basis $\{%
\mathbf{\bar{n}}_{0},\mathbf{\bar{n}}_{1},\mathbf{n}_{2},\mathbf{n}_{3},%
\mathbf{n}_{4}\}$ the structure functions are such that%
\begin{equation}
c_{~4b}^{a}=0=c_{~b4}^{a}~,~~\ \ ~~a\neq 4~.  \label{c}
\end{equation}%
Equations (\ref{dn}, \ref{c}) imply that the exact 4-form in the second term
of Eq.~(\ref{dsF}) is necessarily proportional to ${n^{0}}\wedge {n^{1}}%
\wedge {n}^{2}\wedge {n}^{3}$. So, from the form of ${F}$ in Eq.~(\ref{field}%
), it can be said that the exact 4-form is proportional to ${F}\wedge {F}$.
If the proportionality factor is a constant then the Eq.~(\ref{dsF}) will
become the Maxwell-Chern-Simons (MCS) dynamical equations, or Maxwell
equation in case the proportionality factor be zero. As a conclusion, the
functions ${\mathcal{X}},{\mathcal{Y}}$ in Eq.~(\ref{dsF}) should be
constrained to guarantee a proper dynamics for the electromagnetic field (%
\ref{field}). In particular the flat functions ${\mathcal{X}}_{\text{flat~}},%
{\mathcal{Y}}_{\text{flat~}}$ in Eq. (\ref{flatfunctions}) allows the field (%
\ref{field}) to satisfy Maxwell equations (whatever $X$, $Y$ are). Moreover,
by splitting $\beta $ in ${\mathcal{X}}_{\text{flat~}},{\mathcal{Y}}_{\text{%
flat~}}$into two different values,
\begin{eqnarray}
\mathcal{X}(p) &=&\delta +\varepsilon ~p^{2}+(\beta +\mu _{\mathcal{X}%
}~Q)~p^{-2}~,  \label{XX} \\
\mathcal{Y}(r) &=&\delta +\varepsilon ~r^{2}+(\beta +\mu _{\mathcal{Y}%
}~Q)~r^{-2}~,  \label{YY}
\end{eqnarray}%
then Eq.~(\ref{dsF}) for the field (\ref{field}) becomes a MCS equation:
\begin{equation}
d\ast {F}=-(\mu _{\mathcal{X}}-\mu _{\mathcal{Y}})~{F}\wedge {F~.}
\label{MCS}
\end{equation}

\bigskip

\section{Electrovacuum gravity}

\label{V}

Our aim now is to use the pointlike-charge electromagnetic field (\ref{field}%
) as a source in Einstein equations. The goal will be to solve them by
finding suitable functions $X(p)$, $Y(r)$. The solution must become the $5D$
Kerr-NUT-(A)dS geometry (\ref{KerrNUT}) when the charge $Q$ is zero.

Let us examine the form of Einstein tensor in the $dxKS$ \textit{Ansatz}.
Again we can resort to the structure functions $c_{~bc}^{a}$. By defining $%
c_{abc}=\breve{\eta}_{ad}~c_{bc}^{d}$ , the connection takes the form
\begin{equation}
\Gamma _{bc}^{a}=\frac{1}{2}~\breve{\eta}^{ad}~\left[ c_{cdb}+c_{bdc}-{c}%
_{dbc}\right] ,
\end{equation}%
and the components of the Riemann tensor are
\begin{equation}
{R}_{~~bcd}^{a}=\mathbf{n}_{c}(\Gamma _{bd}^{a})-\mathbf{n}_{d}(\Gamma
_{bc}^{a})+\Gamma _{bd}^{e}~\Gamma _{ec}^{a}-\Gamma _{bc}^{e}~\Gamma
_{ed}^{a}-c_{cd}^{e}~\Gamma _{be}^{a}~,
\end{equation}%
where the vectors $\mathbf{n}_{c}$ act as operators on the functions $\Gamma
_{bd}^{a}$ .

\bigskip

In a general $dxKS$ \textit{Ansatz} the mixed Einstein tensor turns out to
be diagonal only in the block corresponding to the null tetrad. However, if
the functions $\mathcal{X}(p)$ and $\mathcal{Y}(p)$ are restricted to be the
ones of Eqs.~(\ref{XX}) and (\ref{YY}), then the mixed Einstein tensor ${G}$
is entirely diagonal in the basis $\{\mathbf{\bar{n}}_{0},\mathbf{\bar{n}}%
_{1},\mathbf{n}_{2},\mathbf{n}_{3},\mathbf{n}_{4}\}$ of the \textit{Ansatz} (%
\ref{ansatz1}, \ref{ansatz2});\ moreover, ${G}$ is linear in the unknown
functions $X(p)$, $Y(r)$ \cite{DG,GG},
\begin{equation}
{G}=f(r,p)~\left( \mathbf{\bar{n}}_{0}\otimes {n^{0}}+\mathbf{n}_{2}\otimes {%
\underline{n}}^{2}\right) +g(r,p)~\left( \mathbf{\bar{n}}_{1}\otimes {n^{1}}+%
\mathbf{n}_{3}\otimes {\underline{n}}^{3}\right) +h(r,p)~\mathbf{n}%
_{4}\otimes {\underline{n}}^{4}~,  \label{Einstein}
\end{equation}%
where
\begin{eqnarray}
&&f(r,p)=\frac{-p~r~\left( r^{2}-p^{2}\right) ~X^{\prime \prime
}(p)-2~r^{3}~X^{\prime }(p)-\left( p^{3}-3~p~r^{2}\right) ~Y^{\prime }(r)}{%
2~p~r~(r^{2}-p^{2})^{2}}+f_{0}(r,p)  \label{eqf} \\
\notag \\
&&g(r,p)=\frac{p~r~\left( r^{2}-p^{2}\right) ~Y^{\prime \prime
}(r)-2~p^{3}~Y^{\prime }(r)-\left( r^{3}-3~r~p^{2}\right) ~X^{\prime }(p)}{%
2~p~r~(r^{2}-p^{2})^{2}}+g_{0}(r,p)  \label{eqg} \\
\notag \\
&&h(r,p)=-\frac{X^{\prime \prime }(p)-Y^{\prime \prime }(r)}{2~(r^{2}-p^{2})}%
+h_{0}(r,p)  \label{eqh}
\end{eqnarray}%
with
\begin{equation}
f_{0}(r,p)=-g_{0}(r,p)=-\frac{6~p^{4}r^{4}~(\beta +\mu _{\mathcal{X}%
}Q)(\beta +\mu _{\mathcal{Y}}Q)+r^{6}\left( r^{2}-4~p^{2}\right) (\beta +\mu
_{\mathcal{X}}Q)^{2}+p^{6}\left( p^{2}-4~r^{2}\right) (\beta +\mu _{\mathcal{%
Y}}Q)^{2}}{p^{4}~r^{4}\left( r^{2}-p^{2}\right) ^{3}}  \label{f0}
\end{equation}

\medskip

\begin{eqnarray}
h_{0}(r,p)=\frac{6~Q~(\mu _{\mathcal{Y}}-\mu _{\mathcal{X}})\left(
r^{2}(\beta +\mu _{\mathcal{X}}Q)-p^{2}(\beta +\mu _{\mathcal{Y}}Q)\right) }{%
\left( r^{2}-p^{2}\right) ^{4}} &-&\frac{3~p^{2}\left( 2~r^{2}-p^{2}\right)
(\beta +\mu _{\mathcal{Y}}Q)^{2}}{r^{4}~\left( r^{2}-p^{2}\right) ^{3}}
\notag \\
&+&\frac{3~r^{2}\left( 2~p^{2}-r^{2}\right) (\beta +\mu _{\mathcal{X}}Q)^{2}%
}{p^{4}~\left( r^{2}-p^{2}\right) ^{3}}.  \label{h0}
\end{eqnarray}

\bigskip

\subsection{Vacuum solutions}

Einstein vacuum equations with cosmological constant $\Lambda $ are%
\begin{equation}
{G}+\Lambda ~\mathbbm{1}=0~.
\end{equation}%
In the $dxKS$ framework they amount to
\begin{equation}
f(r,p)=g(r,p)=h(r,p)=-\Lambda ~.
\end{equation}%
It is easy to verify that the $5D$ Kerr-NUT-(A)dS geometry (\ref{KerrNUT})
solves these equations ($Q$ must be taken as zero in Eqs.~(\ref{f0}, \ref{h0}%
)). Anyway we will recover this result as a particular case of the charged
geometry to be solved below.

\bigskip

\subsection{Electrovacuum solutions}

Einstein electrovacuum equations are%
\begin{equation}
{G}+\Lambda ~\mathbbm{1}=8\pi ~{T}~,  \label{Einstein-ev}
\end{equation}%
where ${T}$ is the mixed electromagnetic energy-momentum tensor,%
\begin{equation}
T_{~~\nu }^{\mu }=\frac{1}{4\pi }\left( F^{\mu \beta }~F_{\beta \nu }-\frac{1%
}{4}~\delta _{\nu }^{\mu }~F^{\alpha \beta }~F_{\beta \alpha }\right) ~.
\label{tem}
\end{equation}%
This tensor is valid even in the MCS case because the Chern-Simons term in
the Lagrangian, the 5-form ${A\wedge F\wedge F}$, does not contain the
metric;\ thus it does not contribute to the energy-momentum tensor. The
expression of ${T}$\ for the pointlike-charge field (\ref{field}) is
\begin{equation}
{T}=-\frac{Q^{2}}{2\pi \left( r^{2}-p^{2}\right) ^{3}}\left( \mathbf{\bar{n}}%
_{0}\otimes {n}^{0}-\mathbf{\bar{n}}_{1}\otimes {n}^{1}+\mathbf{n}%
_{2}\otimes {\underline{n}}^{2}-\mathbf{n}_{3}\otimes {\underline{n}}%
^{3}\right) -\frac{Q^{2}\left( r^{2}+p^{2}\right) }{2\pi \left(
r^{2}-p^{2}\right) ^{4}}~\mathbf{n}_{4}\otimes {\underline{n}}^{4}~.
\label{tem-pointlike}
\end{equation}%
The $(2+2)D$ sector in ${T}$ is traceless. The additional dimension provides
${T}$ with a non-null trace. These trace will be responsible for preventing
the fulfillment of Maxwell equations, as we are going to show.

By comparing Eqs.~(\ref{Einstein}) and (\ref{tem-pointlike}) one realizes
that Eq.~(\ref{Einstein-ev}) implies three equations:
\begin{eqnarray}
&&f(r,p)+\Lambda -\frac{4~Q^{2}}{(r^{2}-p^{2})^{3}}=0\ ,  \label{eqfem} \\
&&g(r,p)+\Lambda +\frac{4~Q^{2}}{(r^{2}-p^{2})^{3}}=0\ ,  \label{eqgem} \\
&&h(r,p)+\Lambda +\frac{4~Q^{2}~(r^{2}+p^{2})}{(r^{2}-p^{2})^{4}}=0\ .
\label{eqhem}
\end{eqnarray}%
Let us focus on the last one. Combining it with Eq.~(\ref{eqh}) and
multiplying by $2(p^{2}-r^{2})$ we obtain
\begin{equation}
X^{\prime \prime }(p)-Y^{\prime \prime }(r)-2~(r^{2}-p^{2})\left(
h_{0}(r,p)+\Lambda +\frac{4~Q^{2}~(r^{2}+p^{2})}{(r^{2}-p^{2})^{4}}\right)
=0~.  \label{X''Y''}
\end{equation}%
Then, the last term must split into a sum of a function of $r$ and a
function of $p$; so the second derivative $\partial ^{2}/\partial r\partial
p $ of this term must be zero:
\begin{equation}
\frac{\partial ^{2}}{\partial r~\partial p}\left[ 2~(r^{2}-p^{2})\left(
h_{0}(r,p)+\Lambda +\frac{4~Q^{2}~(r^{2}+p^{2})}{(r^{2}-p^{2})^{4}}\right) %
\right] =\frac{96~Q^{2}~p~r~\left( 3(\mu _{\mathcal{Y}}-\mu _{\mathcal{X}%
})^{2}-4\right) ~\left( r^{2}+p^{2}\right) }{(r^{2}-p^{2})^{5}}=0~,
\label{mumu}
\end{equation}%
which implies the necessary condition%
\begin{equation}
(\mu _{\mathcal{Y}}-\mu _{\mathcal{X}})^{2}=\frac{4}{3}~.
\end{equation}%
Thus, the pointlike-charge solution only works if associated with the
specific Chern-Simons coupling constant $|\mu _{\mathcal{Y}}-\mu _{\mathcal{X%
}}|=2/\sqrt{3}$; at least this is the result in the $dxKS$ framework.
Remarkably, the term $4$ in the parenthesis of Eq.~(\ref{mumu}) comes from
the $\mathbf{n}_{4}\otimes {n^{4}}$ term in Eq.~(\ref{tem-pointlike}) (i.e.,
it comes from the trace of ${T}$); on the other hand, the term $3(\mu _{%
\mathcal{Y}}-\mu _{\mathcal{X}})^{2}$ comes from $h_{0}(r,p)$. Thus, the
value of the CS coupling constant is dictated by the the additional
dimensions that provide ${T}$ with a trace.\footnote{%
Nevertheless, solutions to Maxwell equations do exist in five dimensions for
those cases where the energy-momentum tensor results to be traceless. See,
for instance, the pure radiation solution in Section VI of Ref. \cite%
{ferraro}.} Presumably, each odd dimension will correspond to a different CS
coupling constant.

\bigskip

Coming back to Eq.~(\ref{X''Y''}), now we have
\begin{equation}
X^{\prime \prime }(p)-Y^{\prime \prime }(r)+\frac{6~(\beta +\mu _{\mathcal{X}%
}~Q)^{2}}{p^{4}}+2~\Lambda ~p^{2}-\frac{6~(\beta +\mu _{\mathcal{Y}}~Q)^{2}}{%
r^{4}}-2~\Lambda ~r^{2}=0\ .
\end{equation}%
By using the separation constant $2\alpha $ it yields%
\begin{eqnarray}
X^{\prime \prime }(p)+2~\Lambda ~p^{2}+6~(\beta +\mu _{\mathcal{X}%
}~Q)^{2}~p^{-4}-2~\alpha  &=&0\ , \\
Y^{\prime \prime }(r)+2~\Lambda ~r^{2}+6~(\beta +\mu _{\mathcal{Y}%
}~Q)^{2}~r^{-4}-2~\alpha  &=&0\ ,
\end{eqnarray}%
which have solutions
\begin{eqnarray}
X(p) &=&d_{2}~p+d_{1}-\frac{\Lambda }{6}~p^{4}-(\beta +\mu _{\mathcal{X}%
}~Q)^{2}~p^{-2}+\alpha ~p^{2}\ , \\
Y(r) &=&c_{2}~r+c_{1}-\frac{\Lambda }{6}~r^{4}-(\beta +\mu _{\mathcal{Y}%
}~Q)^{2}~r^{-2}+\alpha ~r^{2}\ .
\end{eqnarray}%
Replacing these expressions for $X$ and $Y$, Eqs.~(\ref{eqfem}) and (\ref%
{eqgem}) turn out to be
\begin{eqnarray}
2~c_{2}~r^{3}+d_{2}~\left( p^{3}-3~p~r^{2}\right)  &=&0\ , \\
2~d_{2}~p^{3}+c_{2}~\left( r^{3}-3~r~p^{2}\right)  &=&0\ ,
\end{eqnarray}%
which hold true only if $c_{2}=d_{2}=0$. As in the vacuum case, $c_{1}=-2m$
is interpreted as the mass, and $d_{1}=2n$ as the NUT charge; but only their
difference is physically meaningful since they can be added with a constant $%
\gamma $ without affecting the solution. In sum, the charged rotating black
hole solution to Einstein-Maxwell-Chern-Simons equations (\ref{Einstein-ev}, %
\ref{MCS}) in the $dxKS$ \textit{Ansatz} (\ref{dxKSinverse}, \ref{dxKSansatz}%
) is characterized by the functions
\begin{eqnarray}
X(p)=2n+\gamma +\alpha ~p^{2}-\frac{\Lambda }{6}~p^{4}-(\beta +\mu _{%
\mathcal{X}}~Q)^{2}~p^{-2}\ ,~\ ~~\ \ \ \ ~~\mathcal{X}(p) &=&\delta
+\varepsilon ~p^{2}+(\beta +\mu _{\mathcal{X}}~Q)~p^{-2}~,  \label{solx} \\
&&  \notag \\
Y(r)=-2m+\gamma +\alpha ~r^{2}-\frac{\Lambda }{6}~r^{4}-(\beta +\mu _{%
\mathcal{Y}}~Q)^{2}~r^{-2}\ ,~\ ~~\ \ \ \ ~~\mathcal{Y}(r) &=&\delta
+\varepsilon ~r^{2}+(\beta +\mu _{\mathcal{Y}}~Q)~r^{-2}~,  \label{soly}
\end{eqnarray}%
where $|\mu _{\mathcal{Y}}-\mu
_{\mathcal{X}}|=2/\sqrt{3}$.\footnote{Equations (\ref{MCS},
\ref{f0}, \ref{h0}) show that changing the signs of $\mu
_{\mathcal{X}},\; \mu _{\mathcal{Y}}$ is equivalent to changing the
sign of $Q$; so it is enough to consider that $\mu
_{\mathcal{Y}}-\mu _{\mathcal{X}} =2/\sqrt{3}$ when replacing in the
equations.}

As mentioned in Section \ref{II}.D, the free constants
$\alpha $, $\beta $, $\gamma $, $\delta $, $\varepsilon $ represent
choices of charts in the maximally symmetric case ($m+n=0=Q$).
Clearly, $\gamma$ and $\beta$  in the solution  (\ref{solx},
\ref{soly}) are absorbed in a shift of $m$, $n$, $ \mu
_{\mathcal{X}}$,  $ \mu _{\mathcal{Y}}$,
\begin{equation}
m\longrightarrow m'=m-\frac{\gamma}{2}\ , \  \  \  \  \ n\longrightarrow n'=n
+\frac{\gamma}{2}\ , \  \  \  \  \   \mu _{\mathcal{X,Y}}\longrightarrow
\mu '_{\mathcal{X,Y}}= \mu _{\mathcal{X,Y}}+\beta\; Q^{-1}\  ,\label{shift}
\end{equation}
the magnitudes $m+n$ and $ \mu _{\mathcal{X}}-\mu _{\mathcal{Y}}$
being unaffected by the shift.
The CCLP solution as originally made known \cite{cclp, DKL}, was built with
$\mu' _{\mathcal{X}}=0$ and $\mu' _{\mathcal{Y}}=2/\sqrt{3}$. By
choosing $\delta=0=\epsilon$ one
concludes that the CCLP solution fits the $xKS$ form with the (A)dS metric
playing the role of $\mathbf{h}$ \cite{Aliev, xKS}.\footnote{A
suitable coordinate transformation is needed to compare the components of
the metric with those of Refs.~\cite{cclp,ferraro} (see Appendix \ref{apen}).}

The shift (\ref{shift}) could suggest that the only
physically meaningful magnitudes are  $m+n$ and $ \mu
_{\mathcal{Y}}-\mu _{\mathcal{Y}}$, so extending the degeneration
between $m$ and $n$ typical of the uncharged solution. However, as
shown in the Appendix \ref{apenK}, the form of the Kretschmann
scalar for the charged solution indicates that such degeneration is
actually broken by the charge.

\section{Signature $(1,4)$}

\label{VI}

Up to now we have worked with a metric of signature $(3,2)$, which greatly
simplifies the algebraic aspects of the problem by allowing the metric to
take the \textit{dxKS} form. Even though the signature $(3,2)$ can be
relevant on its own in different contexts (two-time physics \cite{bars},
AdS/dS spacetimes \cite{ads1,ads2}, Kleinian spaces \cite{klein}), an
important feature of our approach is that we can retrieve the Lorentzian
signature by means of a proper Wick rotation.

Let us consider the coordinate transformation
\begin{eqnarray}
&&dt^{\prime }=dt-\frac{a^{2}+b^{2}+p^{2}+ab~\mathcal{X}(p)}{X(p)}~dp-\frac{%
a^{2}+b^{2}+r^{2}+ab~\mathcal{Y}(r)}{Y(r)}~dr~,  \label{coc1} \\
&&  \notag \\
&&d\phi ^{\prime }=d\phi -\frac{a+b~\mathcal{X}(p)}{\left(
a^{2}-b^{2}\right) ~X(p)}a^{2}dp-\frac{a+b~\mathcal{Y}(r)}{\left(
a^{2}-b^{2}\right) ~Y(r)}a^{2}dr~,  \label{coc2} \\
&&  \notag \\
&&d\psi ^{\prime }=d\psi +\frac{b+a~\mathcal{X}(p)}{\left(
a^{2}-b^{2}\right) ~X(p)}b^{2}dp+\frac{b+a~\mathcal{Y}(r)}{\left(
a^{2}-b^{2}\right) ~Y(r)}b^{2}dr~,  \label{coc3}
\end{eqnarray}%
The components of the metric (\ref{dxKSg}) in the chart $(t^{\prime },\phi
^{\prime },\psi ^{\prime },r,p)$ are%
\begin{equation}
g_{\mu \nu }=\left(
\begin{array}{ccc}
\mathbf{M} & 0 & 0 \\
0 & \frac{-p^{2}+r^{2}}{Y(r)} & 0 \\
0 & 0 & \frac{-p^{2}+r^{2}}{X(p)}%
\end{array}%
\right) ~.  \label{metricnew}
\end{equation}%
So the block $r,p$ has been diagonalized; instead, $\mathbf{M}$ remains as a
non-diagonal $3\times 3$ matrix. In particular the vectors $\partial
/\partial r$ and $\partial /\partial p$ are no longer null in this chart.%
\footnote{%
Vectors $\partial /\partial r$, $\partial /\partial p$ has changed, since
they now differentiate along curves of constant $t^{\prime },\phi ^{\prime
},\psi ^{\prime }$.} The coordinate transformation (\ref{coc1})-(\ref{coc3})
preserves the volume,
\begin{equation}
\det (g_{\mu \nu })=\frac{r^{2}p^{2}}{a^{4}b^{4}}%
~(a^{2}-b^{2})^{2}~(-p^{2}+r^{2})^{2}~.
\end{equation}

By analyzing the form of the components $g_{\mu \nu }$ in Eq.~(\ref%
{metricnew}), one realizes that the metric tensor can be displayed in the
extended Pleba\'{n}ski-Demia\'{n}ski \textit{Ansatz }of Ref.~\cite{ferraro},%
\footnote{%
The exact comparison with Ref. \cite{ferraro} is provided in
Appendix \ref{apen}.}
\begin{equation}
\mathbf{g}=-\frac{Y(r)}{-p^{2}+r^{2}}~\mathbf{\omega }^{0}\otimes \mathbf{%
\omega }^{0}+\frac{X(p)}{-p^{2}+r^{2}}~\mathbf{\omega }^{1}\otimes \mathbf{%
\omega }^{1}+\frac{r^{2}p^{2}}{a^{2}b^{2}}~\mathbf{\Omega }^{2}\otimes
\mathbf{\Omega }^{2}+\frac{-p^{2}+r^{2}}{Y(r)}~\mathbf{\omega }^{3}\otimes
\mathbf{\omega }^{3}-\frac{-p^{2}+r^{2}}{X(p)}~\mathbf{\omega }^{4}\otimes
\mathbf{\omega }^{4}~,  \label{PD}
\end{equation}%
where the 1-forms $\mathbf{\omega }^{a},\mathbf{\Omega }^{2}$ in the
coordinate basis $\{dt^{\prime },d\phi ^{\prime },d\psi ^{\prime },dr,dp\}$
read (after dropping the primes)
\begin{eqnarray}
\mathbf{\omega }^{0} &=&dt-a^{-1}(a^{2}+p^{2})~d\phi
-b^{-1}(b^{2}+p^{2})~d\psi  \label{o1} \\
\mathbf{\omega }^{1} &=&dt-a^{-1}(a^{2}+r^{2})~d\phi
-b^{-1}(b^{2}+r^{2})~d\psi  \label{o2} \\
\mathbf{\omega }^{2} &=&\frac{b^{2}}{a}~d\phi +\frac{a^{2}}{b}~d\psi ~,~~~\
\ ~\ \ ~~~~~~\mathbf{\Omega }^{2}=\mathbf{\omega }^{2}-\frac{a~b}{p^{2}-r^{2}%
}~(\mathcal{Y}(r)~\mathbf{\omega }^{0}-\mathcal{X}(p)~\mathbf{\omega }^{1})
\label{o3} \\
\mathbf{\omega }^{3} &=&dr~,~~~~~~~~~\ \ ~\ ~~~\mathbf{\omega }^{4}=dp~.
\label{o4}
\end{eqnarray}%
As can be seen , the chart we are using here is simpler than the one used in
Ref.~\cite{ferraro}.

\medskip

Notice that the coordinate transformation (\ref{coc1})-(\ref{coc3}) involves
the functions $X,Y,\mathcal{X},$ $\mathcal{Y}$ carrying the information on
the gravitational field. So the components $g_{\mu \nu }$ in the new chart
can no longer be split into the sum of the components of a flat seed metric
plus gravitational terms. Namely, the choice $X=0=Y$ as a way to recover the
flat metric is not longer allowed.

\bigskip

In Section \ref{II} we have started with a flat seed metric of signature $%
(3,2)$; equation (\ref{PD}) shows that this signature is kept for every
choice of $X,Y$ and every sign of $r^{2}-p^{2}$. Let us perform the Wick
rotation of $p$,
\begin{equation}
p\longrightarrow i~p~,
\end{equation}%
therefore%
\begin{equation}
\mathbf{\omega }^{4}\otimes \mathbf{\omega }^{4}\longrightarrow -\mathbf{%
\omega }^{4}\otimes \mathbf{\omega }^{4}~.
\end{equation}%
If we further assume that that the functions $X(ip)$, $\mathcal{X}(ip)$ are
real --which is indeed the case for the solutions in Eqs.~(\ref{solx}, \ref%
{soly})--, it follows that the metric signature depends only on the sign of $%
X(ip)$: if $X(ip)<0$, then the (real) metric (\ref{PD}) gets a Lorentzian
signature $(1,4)$. The Wick rotation, besides, puts on an equal footing the
coordinates $r$, $p$, since the ubiquitous expression $(-p^{2}+r^{2})$
becomes $(p^{2}+r^{2})$. The zenith angle $\theta$
of $5D$ ellipsoidal coordinates is introduced through the
relation $p^2=a^2\cos^2\theta+b^2\sin^2\theta$.

\medskip

It is important to note that although not every metric in the \emph{dxKS}
ansatz is transformed to the signature $(1,4)$ by the proposed coordinate
transformation (if $X(ip)>0$ the Wick rotation does not change the
signature), every Lorentzian metric that can be expressed in the Pleba\'{n}%
ski-Demia\'{n}ski \textit{Ansatz} can also be expressed in the \emph{dxKS}
form by performing the inverse transformation of (\ref{coc1})-(\ref{coc3}),
together with the corresponding Wick rotation.

\section{Conclusions}

\label{VII}

Equations (\ref{solx}, \ref{soly}) display the ``pointlike''
rotating electrovacuum solution in the $5D$ double extended
Kerr-Schild ($dxKS$) framework (\ref{dxKSinverse},
\ref{dxKSansatz}). The physical parameters in Eqs.~(\ref{solx},
\ref{soly}) are the mass $m$, the NUT charge $n$, $\mu
_{\mathcal{X}}Q$, and $\mu _{\mathcal{Y}}Q$. The difference $(\mu
_{\mathcal{Y}}-\mu _{\mathcal{X}})Q$, which connects with the
Chern-Simons coupling constant in Eq.~(\ref{MCS}), is $2Q/\sqrt{3}$,
as dictated by the trace of the Einstein equations. The degeneracy
between $m$ and $n$ exhibited by the uncharged solution is broken by
the presence of the electric charge $Q$, as evidenced by the
Kretschmann scalar. Since the EMCS equations depend on
$\mu_\mathcal{X}$ and $\mu_\mathcal{Y}$ only through their
difference, one might expect a degeneracy between these parameters.
However, by inspecting the Kretschmann scalar one concludes this is
not the case. As the Ricci tensor depends only on
$\mu_\mathcal{X}-\mu_\mathcal{Y}$, the degeneracy breaking is
located in the Weyl tensor.

\bigskip

Let us review the basic elements of the $5D$ $dxKS$ \textit{Ansatz}. We
start from a flat $(2+3)D$ seed spacetime, whose tangent and cotangent
spaces are spanned by the dual bases ${\{}\mathbf{n}_{a}\}$ and ${\{n}^{a}\}$
respectively (see Eq.~(\ref{inverse})). The (by definition) null tetrad $\{%
\mathbf{n}_{0},~\mathbf{n}_{1},~\mathbf{n}_{2}\doteq \mathbf{k},~\mathbf{n}%
_{3}\doteq \mathbf{K}\}$ is constituted by four independent geodesic
congruences. The way the \textit{Ansatz} (\ref{ansatz1}, \ref{ansatz2})
introduces gravity implies the deformation of $\mathbf{n}_{0},~\mathbf{n}%
_{1} $, by following some rules that are essential for the pointlike
electromagnetic potential (\ref{potential}) to work:

\medskip

i) ${n}^{0}\wedge {n}^{2}$ and ${n}^{1}\wedge {n}^{3}$ do not change.

\medskip

ii) the spacetime volume is preserved (the determinant of the metric is not
affected).

\medskip

iii) $\mathbf{k}$,$~\mathbf{K}$ continue to be null and geodesic vectors in
the deformed metric.

\medskip

iv) Since the \textit{Ansatz} does not modify the vectors $\mathbf{k}$, $%
\mathbf{K}$, they are Lie-dragged by the independent Killing vectors $%
\partial _{t}$, $\partial _{\phi }$, $\partial _{\psi }$; in fact they are
still the coordinate vectors $\mathbf{k}=\partial /\partial r$, $\mathbf{K}%
=\partial /\partial p$. In particular, the $5D$ $dxKS$ \textit{Ansatz} keeps
the two independent \textquotedblleft rotational\textquotedblright\
symmetries of the seed metric.

\medskip

v) The \textit{Ansatz} protects the property (\ref{c}). This property means
that $d{n}^{a}(\mathbf{m})=0$ if $a\neq 4$; Eq.~(\ref{dkdK}) corresponds to
the cases $a=0,1$. \medskip

\medskip

One might consider a further extension of \textit{Ansatz} (\ref{dxKSinverse}) by adding a term proportional to $\mathbf{k}\otimes\mathbf{K}+\mathbf{K}\otimes\mathbf{k}$. It turns out that, although such metric preserves most of the properties posed by (\ref{dxKSinverse}), by solving the EMCS equations under this assumption we observe that they can only be fulfilled when this term is absent; at least for the electromagnetic source in Eq. (\ref{potential}).

\medskip

Although we develop the $5D$ $dxKS$ \textit{Ansatz} in the context of metric
signature $(3,2)$, in Section \ref{VI} we have displayed a change of
coordinate followed by a Wick rotation that throws the metric into the
signature $(1,4)$, and leaves it in the Pleba\'{n}ski-Demia\'{n}ski \textit{%
Ansatz} of Ref.~\cite{ferraro}.

The computational simplicity is an important issue in this type of work. The
charts we introduced in Sections \ref{II} and \ref{VI}, which make the bases
look as shown in Eqs.~(\ref{n2n3}- \ref{mu}) and ~(\ref{o1}- \ref{o4}),
greatly facilitate the computational procedures, so providing a new way to
advance in the search for higher-dimensional electrovacuum solutions.

\bigskip

\begin{acknowledgments}
This work was supported by Consejo Nacional de Investigaciones
Cient\'{\i}ficas y T\'{e}cnicas (CONICET) and Universidad de Buenos Aires.
\end{acknowledgments}

\section*{Data availability statement}

Data sharing is not applicable to this article as no datasets were generated or analysed during the current study.

\appendix

\section{Kretschmann scalar}\label{apenK}

The Kretschmann scalar for the solution (\ref{solx}, \ref{soly}) is
\begin{eqnarray}
K &=&\frac{10~\Lambda ^{2}}{9}+96~(m+n)^{2}\  \frac{
3p^{4}+10p^{2}r^{2}+3r^{4}}{\left( p^{2}-r^{2}\right)^{6}}- \frac{768 Q^2 (\mu _{\mathcal{X}}\!-\!\mu
_{\mathcal{Y}})}{\left(p^2-r^2\right)^6}\;(m\!+\!n)  (\mu' _{\mathcal{X}}\!+\!\mu'
_{\mathcal{Y}}) \;(p^2+r^2)\notag  \\  \notag  \\
&&+\frac{4\Lambda Q^{2}(\mu _{\mathcal{X}}\!-\!\mu
_{\mathcal{Y}})^{2}}{3\left(
p^{2}-r^{2}\right) ^{6}}\left( p^{6}\!+23p^{4}r^{2}\!+23p^{2}r^{4}\!+r^{6}\right)\notag  \\  \notag  \\
&& +
\frac{48Q^{2}(\mu _{\mathcal{X}}\!-\!\mu _{\mathcal{Y}})^{2}}{\left(
p^{2}-r^{2}\right) ^{7}}
\left((m\!+\! n) (p^{4}\! +\! 26 p^2 r^2\! +\! r^4)-8(m' \; p^{4}\!+\! n' \; r^{4})\right)
\notag  \\  \notag  \\
&&-\frac{4 Q^4 (\mu
_{\mathcal{X}}\!-\!\mu
_{\mathcal{Y}})^2}{\left(-p^2+r^2\right)^8} \left((\mu _{\mathcal{X}}\!-\!\mu
_{\mathcal{Y}})^2\left(28 p^2 r^2+65\left(p^4+r^4\right)\right)-192 (\mu'_{\mathcal{X}}\; p^2
- \mu'_{\mathcal{Y}}\; r^2 )^2\right)\ .
\end{eqnarray}
Unlike the scalars $R$ and $R_{ab}R^{ab}$,  $K$ depends not only on $m+n$, $\mu _{\mathcal{X}}\!-\!\mu
_{\mathcal{Y}}$, but on $m'$, $n'$,  $\mu' _{\mathcal{X}}$,  $\mu' _{\mathcal{Y}}$ in an independent way.
The terms breaking the degeneration are linked to the electric charge $Q$. Since the Kretschmann
scalar admits the decomposition
\begin{equation}
K=R_{abcd}R^{abcd}=C_{abcd}C^{abcd}+\frac{4}{d-2}R_{ab}R^{ab}-\frac{2}{(d-1)(d-2)}R^2 \ ,
\end{equation}
then the contributions breaking the degeneration are located in the Weyl tensor $C_{abcd}$.

The physical and geometric interpretation of the parameters $\gamma$ and $\beta$ in the charged case is yet to be understood. A possibility one must consider is that these constants represent mere coordinate choices. In such case one would be able to absorb these parameters present in the Kretschmann scalar in a redefinition of the coordinates $r$ and $p$. Note, however, that K is linear in $\gamma$ and quadratic in $\beta$, while the coordinates $r$ and $p$ appear in a variety of powers ranging from $-12$ to $0$. Furthermore, once the coordinate change is performed the scalars $R$ and $R^{ab}R_{ab}$ should continue to be independent of $\gamma$ and $\beta$. Because of these reasons we can see that such a change of coordinates would be hard to obtain if not impossible.

If the parameters $\gamma$ and $\beta$ cannot be absorbed in a coordinate change then different values of these parameters would represent different spacetimes. The fact that these constants appear on the Weyl tensor seems to indicate that they might play a relevant role in the algebraic classification of these spacetimes \cite{coley}.

\section{Relationship with the Pleba\'{n}ski-Demia\'{n}ski \textit{Ansatz}
in Ref.~\protect\cite{ferraro}} \label{apen}
Equation~(\ref{PD}) in Section \ref{VI} displays an extended Pleba\'{n}%
ski-Demia\'{n}ski \textit{Ansatz} that works for signature $(3,2)$; it uses
a much simpler set of 1-forms $\{\omega ^{a},\mathbf{\Omega }^{2}\}$ than
the one used in Ref.~\cite{ferraro}. For the sake of completeness, we will
show the way to obtain the metric as displayed in Ref.~\cite{ferraro}. Let
us perform the complex coordinate transformation
\begin{eqnarray}
dt^{\prime } &=&dt-\frac{a^{2}+b^{2}+p^{2}+ab~\mathcal{X}(p)}{X(p)}~dp-\frac{%
a^{2}+b^{2}+r^{2}+ab~\mathcal{Y}(r)}{Y(r)}~dr~,  \label{acoc1} \\
&&  \notag \\
d\phi ^{\prime } &=&a~\lambda ~dt+\frac{(1-a^{2}\lambda )(a^{2}-b^{2})}{a^{2}%
}~d\phi -\frac{a(1+\lambda (b^{2}+p^{2}))+b~\mathcal{X}(p)}{X(p)}a^{2}dp-%
\frac{a(1+\lambda (b^{2}+r^{2}))+b~\mathcal{Y}(r)}{Y(r)}a^{2}dr~,
\label{acoc2} \\
&&  \notag \\
d\psi ^{\prime } &=&b~\lambda ~dt-\frac{(a^{2}-b^{2})(1-b^{2}\lambda )}{b^{2}%
}~d\psi +\frac{b(1+\lambda (a^{2}+p^{2}))+a~\mathcal{X}(p)}{X(p)}b^{2}dp+%
\frac{b(1+\lambda (a^{2}+r^{2}))+a~\mathcal{Y}(r)}{Y(r)}b^{2}dr~,
\label{acoc3} \\
&&  \notag \\
dr^{\prime } &=&dr~,~~~~~~~~~\ \ ~\ ~~~dp^{\prime }=-i~dp~.  \label{acoc4}
\end{eqnarray}%
where $\lambda =\Lambda /6$, then we can see that the metric tensor (\ref{dxKSg})
can be expressed as (we drop the primes)
\begin{equation}
\mathbf{g}=-\frac{Y(r)}{p^{2}+r^{2}}~\mathbf{\omega }^{0}\otimes \mathbf{%
\omega }^{0}+\frac{X(ip)}{p^{2}+r^{2}}~\mathbf{\omega }^{1}\otimes \mathbf{%
\omega }^{1}-\frac{a^{2}b^{2}}{r^{2}p^{2}}~\mathbf{\Omega }^{2}\otimes
\mathbf{\Omega }^{2}+\frac{p^{2}+r^{2}}{Y(r)}~dr\otimes dr+\frac{p^{2}+r^{2}%
}{X(ip)}~dp\otimes dp~,  \label{APD}
\end{equation}%
where the 1-forms $\mathbf{\omega }^{a},\mathbf{\Omega }^{2}$ read
\begin{eqnarray}
\mathbf{\omega }^{0} &\equiv &\frac{(1-p^{2}\lambda )~dt}{(1-a^{2}~\lambda
)(1-b^{2}~\lambda )}-\frac{a~(a^{2}-p^{2})~d\phi ^{\prime }}{%
(a^{2}-b^{2})(1-a^{2}~\lambda )}-\frac{b~(b^{2}-p^{2})~d\psi }{%
(b^{2}-a^{2})(1-b^{2}~\lambda )}~,  \label{omega0} \\
&&  \notag \\
\mathbf{\omega }^{1} &\equiv &\frac{(1+r^{2}\lambda )~dt}{(1-a^{2}~\lambda
)(1-b^{2}~\lambda )}-\frac{a~(a^{2}+r^{2})~d\phi }{(a^{2}-b^{2})(1-a^{2}~%
\lambda )}-\frac{b~(b^{2}+r^{2})~d\psi }{(b^{2}-a^{2})(1-b^{2}~\lambda )}~,
\label{omega1} \\
&&  \notag \\
\mathbf{\omega }^{2} &\equiv &-\frac{(1+r^{2}\lambda )~(1-p^{2}\lambda )~dt}{%
(1-a^{2}~\lambda )(1-b^{2}~\lambda )}+\frac{(a^{2}+r^{2})(a^{2}-p^{2})~d\phi
}{a~(a^{2}-b^{2})(1-a^{2}~\lambda )}+\frac{(b^{2}+r^{2})(b^{2}-p^{2})~d\psi
}{b~(b^{2}-a^{2})(1-b^{2}~\lambda )}~  \label{omega2} \\
&&  \notag \\
\mathbf{\Omega }^{2} &\equiv &\mathbf{\omega }^{2}-\frac{p^{2}r^{2}}{%
ab\left( r^{2}+p^{2}\right) }\left[ \left( \mathcal{Y}(r)-\frac{ab}{r^{2}}%
\right) \mathbf{\omega }^{0}-\left( \mathcal{X}(ip)+\frac{ab}{p^{2}}\right)
\mathbf{\omega }^{1}\right] .  \label{Omega2}
\end{eqnarray}%
The metric in Ref.~\cite{ferraro} is retrieved by redefining
$-X(ip)\longrightarrow X(p)$, $-Y(r)\longrightarrow Y(r)$,
$\mathcal{X}(ip)+a b p^{-2}\longrightarrow\mathcal{X}(p)$,
$\mathcal{Y}(r)-a b r^{-2}\longrightarrow\mathcal{Y}(r)$ together
with a global change of sign (Ref. \cite{ferraro} uses signature
$(4,1)$ instead of the signature $(1,4)$ used in this work). As
mentioned in Section \ref{VI}, the metric signature of this ansatz
depends solely on the sign of the
function $X$. If $X(ip)>0$, then the metric (\ref{APD}) has signature $(3,2)$%
. If $X(ip)<0$, the signature is $(1,4)$ and the complex coordinate
transformation (\ref{acoc1})-(\ref{acoc3}) renders the metric (\ref{APD})
Lorentzian.

\end{document}